\documentclass[
    aps,
    prd,
    amssymb,
    superscriptaddress,
    floatfix,
    nofootinbib,
    reprint
]{revtex4-2}

\usepackage{amsmath}
\usepackage{amsfonts}
\usepackage{bm}
\usepackage{braket}
\usepackage{graphicx}
\usepackage{booktabs}
\usepackage{multirow}
\usepackage{dcolumn}
\usepackage{enumitem}
\usepackage{lineno}
\usepackage[dvipsnames,x11names]{xcolor}
\usepackage[normalem]{ulem}
\usepackage{soul}
\usepackage{hyperref}
\hypersetup{
    colorlinks=true,
    citecolor=Green,
    urlcolor=Blue,
    linkcolor=Blue
}
\usepackage{comment}
\usepackage{mfirstuc}
\usepackage{orcidlink}

\usepackage[T1]{fontenc}
\usepackage{microtype}
\usepackage[vvarbb,smallerops,lining]{newtx}

\usepackage[compact]{titlesec}

\titleformat{\section}
    {\centering\small\bfseries\MakeUppercase}
    {\noindent\thesection.}
    {2pt}{}

\titleformat{\subsection}
    {\centering\small\bfseries}
    {\noindent\thesubsection.}
    {2pt}{}

\titlespacing\section{0pt}{18pt plus 1pt minus 1pt}{8pt plus 1pt minus 1pt}
\titlespacing\subsection{0pt}{12pt plus 1pt minus 1pt}{8pt plus 1pt minus 1pt}

\newcommand{\tstrut}{\rule{0pt}{2.6ex}}

\makeatletter

\newdimen\tov@rt
\newcommand{\tovgap}{2}

\newcommand{\tov@setrule}[1]{%
    \ifx#1\scriptscriptstyle
        \tov@rt=\fontdimen8\scriptscriptfont3
    \else\ifx#1\scriptstyle
        \tov@rt=\fontdimen8\scriptfont3
    \else
        \tov@rt=\fontdimen8\textfont3
    \fi\fi
}

\newcommand{\tov@build}[2]{%
    \tov@setrule{#1}
    \sbox0{$\m@th#1#2$}%
    \vbox{%
        \offinterlineskip
        \kern\tov@rt
        \hrule height\tov@rt
        \kern\tovgap\tov@rt
        \box0
    }
}

\newcommand{\tightoverline}[1]{\mathpalette\tov@build{#1}}

\makeatother

\newcommand{\olsi}[1]{\,\tightoverline{\!#1}}

\begin{document}

 \title{\boldmath Coulomb Effects in Momentum-Space Femtoscopy: A Case Study of the $\olsi{K}\Omega$ System}

\newcommand{\ific}{\affiliation{\small%
Instituto de F\'isica Corpuscular, Centro Mixto Universidad de Valencia-CSIC, \\
Institutos de Investigaci\'on de Paterna, Apartado 22085, E-46071, Valencia, Spain}}

\newcommand{\teoUV}{\affiliation{\small%
Departamento de F\'\i sica Te\'orica and IFIC, Centro Mixto Universidad de Valencia-CSIC,
Institutos de Investigaci\'on de Paterna, Aptdo. 22085, E-46071 Valencia, Spain}}

\newcommand{\INFNCatania}{\affiliation{\small%
Istituto Nazionale di Fisica Nucleare, Sezione di Catania, Dipartimento di Fisica ``Ettore Majorana'', Universit\`a di Catania, Via Santa Sofia 64, I-95123 Catania, Italy}}

\author{Pablo Encarnación\orcidlink{0009-0005-0749-3885}}
\email{Pablo.Encarnacion@ific.uv.es}
\teoUV

\author{Amador García-Lorenzo\orcidlink{0009-0001-3200-9996}}
\email{Amador.Garcia@ific.uv.es}
\ific

\author{Miguel Albaladejo\orcidlink{0000-0001-7340-9235}}
\email{Miguel.Albaladejo@ific.uv.es}
\ific 

\author{Albert Feijoo\orcidlink{0000-0002-8580-802X}}
\email{Eduardo.Feijoo@ific.uv.es}
\teoUV

\author{Juan Nieves\orcidlink{0000-0002-2518-4606}}
\email{Juan.M.Nieves@ific.uv.es}
\ific

\author{Isaac Vida\~na\orcidlink{0000-0001-7930-9112}}
\email{isaac.vidana@ct.infn.it}
\INFNCatania

\renewcommand{\abstractname}{\vspace{20pt}Abstract}

\begin{abstract}
We present a momentum-space framework for the consistent treatment of Coulomb interactions in femtoscopic correlation functions based on a modified Vincent--Phatak method that is more amenable to numerical implementation. The formalism provides a practical approach to incorporating Coulomb effects at the short distances relevant for femtoscopy within the Lippmann--Schwinger equation, while preserving a unified treatment of the strong interaction. As an application, we study the $S=-4$ pseudoscalar--baryon decuplet interaction in the $\olsi{K}\Omega$ system and present predictions for the singly and doubly negatively charged channels, $\olsi{K}{}^0\Omega^-$ and $K^-\Omega^-$. As an additional validation of the formalism, we have also applied it to the well-studied $pp$ system. We further assess the limitations of the asymptotic wave-function approximation and quantify corrections accounting for the short-distance structure of the interaction potential. We introduce a phenomenological parameter that effectively absorbs contributions from both the finite source size and the off-shell structure of the interaction, the latter being one of the main obstacles to extracting detailed information on hadron--hadron interactions from femtoscopic measurements in a model-independent way.
\end{abstract}

\maketitle

\setcounter{tocdepth}{3}

\section{Introduction}\label{sec:introduction}

The interaction between mesons and $\Omega^-$ baryons has been scarcely explored so far, mainly because the technical limitations of using such short-lived particles as targets or projectiles preclude the collection of experimental scattering data.

Within the framework of unitarized effective field theories (UEFTs), an early study addressed this problem by deriving the leading $S$-wave interaction of Goldstone bosons with baryon-decuplet states using a chiral coupled-channel approach driven by the Weinberg--Tomozawa (WT) term~\cite{Kolomeitsev:2003kt}. This work was soon extended to several strangeness sectors, including $S=-2$, $-3$, and $-4$, thereby incorporating channels with $\Omega^-$ baryons~\cite{Sarkar:2004jh}. A common outcome of these analyses performed in the early 2000s is the dynamical generation of a $J^P=3/2^-$ resonance, later associated with the $\Omega(2012)^-$, which had not yet been observed at that time. Using the same coupled-channel basis in the $S=-3$ sector, Ref.~\cite{Xu:2015bpl} predicted a bound $J^P=3/2^-$ $\Omega^*$ state near $1800~\text{MeV}$, supporting an alternative low-lying spectrum also favored by an extended quark model with Nambu--Jona--Lasinio interactions~\cite{An:2014lga}. Similarly, attractive next-to-leading-order chiral corrections to the $\olsi{K}\Xi$ interaction lead to a bound $J^P=1/2^-$ state around $1800~\text{MeV}$, together with a $P_{sss}$ pentaquark~\cite{Feijoo:2024qgq}.
Beyond pseudoscalar Goldstone bosons, Ref.~\cite{Sarkar:2010saz} investigated vector-meson--decuplet-baryon interactions within the hidden-gauge formalism~\cite{Bando:1984ej,Bando:1987br,Meissner:1987ge,Nagahiro:2008cv}. An extension of this framework, which includes the $\olsi{K}\Xi^*$ and $\olsi{K}{}^*\Xi$ channels through box-diagram contributions, dynamically generates a state associated with the $\Omega(2380)$ resonance, with properties consistent with current experimental data~\cite{Li:2026wck,Navas:2024aex}.

A more comprehensive treatment was presented in Ref.~\cite{Gamermann:2011mq}, where an $SU(6)$ spin--flavor extension of the meson--baryon chiral WT Lagrangian~\cite{Garcia-Recio:2005elc} was implemented in a coupled-channel unitary framework. Including pseudoscalar and vector mesons interacting with octet and decuplet baryons led to several predicted $\Omega^*$ states, two of them compatible with the observed $\Omega(2250)$ and $\Omega(2380)$ resonances, reinforcing the applicability of such approaches in the exotic $\Omega$ sector.

Strong experimental support for this framework came from the observation of the $\Omega(2012)^-$ by the Belle Collaboration in $e^+e^-$ collisions~\cite{Belle:2018mqs}, later confirmed by BESIII, which also reported indications of a new $\Omega(2109)^-$ structure~\cite{BESIII:2024eqk}, and from subsequent Belle analyses~\cite{Belle:2019zco,Belle:2022mrg}. In particular, the precise determination of the branching-fraction ratio $\mathcal{R}^{\Xi\olsi{K}\pi}_{\Xi\olsi{K}}$~\cite{Belle:2022mrg} provides a stringent test for models of excited $\Omega$ states. Owing to its proximity to the $\olsi{K}\Xi(1530)$ threshold, the $\Omega(2012)^-$ has been widely interpreted as a dynamically generated hadronic molecule in coupled-channel approaches~\cite{Valderrama:2018bmv,Lin:2018nqd,Pavao:2018xub,Huang:2018wth,Gutsche:2019eoh,Lu:2020ste,Ikeno:2020vqv,Lin:2019tex,Ikeno:2022jpe,Lu:2022puv,Han:2025gkp,Shen:2025xcq}, consistently reproducing the measured branching ratio. Furthermore, the ALICE Collaboration has reported a structure near $2013~\text{MeV}$ in the $K^0_S\Xi^-$ invariant-mass spectrum from $pp$ collisions, compatible with the $\Omega(2012)^-$ state~\cite{ALICE:2025atb}.

In addition to this progress, femtoscopy provides a complementary probe of hadron--$\Omega^-$ interactions through momentum correlations that are sensitive to final-state interactions. The previous experimental indications, together with measurements of the $p\Omega^-$ correlation function (CF)~\cite{STAR:2018uho,ALICE:2020mfd} and subsequent theoretical studies~\cite{Morita:2016auo,Morita:2019rph,Piscitelli:2025zxs,iMendez:2024eyq}, has stimulated considerable interest in hadron--$\Omega^-$ interactions. The quark delocalization color screening model indicates that the attractive $\phi\Omega^-$ interaction in the $J^P=1/2^-$ channel is not strong enough to generate a state, resulting in essentially featureless CFs~\cite{Yan:2026yrd}. In contrast, Refs.~\cite{Lin:2026ypf,Liu:2026zlk} predicted the $\olsi{K}{}^0\Xi^{*-}$, $K^-\Xi^{*0}$, and $\eta\Omega^-$ CFs in a framework supporting the molecular interpretation of the $\Omega(2012)^-$~\cite{Pavao:2018xub}. The resulting $\olsi{K}\Xi^*$ CFs exhibit clear signatures of the $\Omega(2012)^-$, providing a benchmark for future femtoscopic measurements.

With our focus on meson--$\Omega^-$ interactions within UEFTs combined with the femtoscopy formalism and techniques, particular attention must be paid to the interplay between strong and Coulomb interactions owing to the charged nature of $\Omega^-$. A reliable scattering wave function, a key ingredient of the CF, requires an accurate treatment of Coulomb effects, which becomes especially relevant at low relative momenta.

The simultaneous treatment of Coulomb and strong interactions is nontrivial due to their different theoretical formulations. The Coulomb force is long-ranged and naturally expressed in coordinate space, whereas the strong interaction derived from EFT is formulated in momentum space and constrained only on-shell. Its off-shell extension, often implemented through separable representations, leads to a nonlocal coordinate-space structure, which prevents a straightforward solution of the Schr\"odinger equation  and restricts analytic treatments to simplified cases that fail to reproduce realistic EFT interactions. In practical applications, the Schr\"odinger equation is solved numerically, but the non-locality of EFT-based interactions makes this procedure technically demanding and typically requires approximations. As a result, local potential models have been widely employed. For instance, in early hadron femtoscopy studies, Koonin~\cite{Koonin:1977fh} evaluated the $pp$ CF by solving the Schr\"odinger equation with a Coulomb potential and a phenomenological strong interaction including central, spin-orbit, and tensor local terms~\cite{Reid:1968sq}. More recently, the ALICE Collaboration has successfully compared its measurements of the $pp$ correlation function~\cite{ALICE:2019buq} with calculations based on a combined Coulomb plus Argonne V18 local potential~\cite{Wiringa:1994wb}, including $S$-, $P$-, and $D$-waves. Local coupled-channel $\olsi{K}N$--$\pi\Sigma$--$\pi\Lambda$ potentials~\cite{Miyahara:2018onh} have been also used to compute the scattering wave functions entering the $K^-p$ CF~\cite{Kamiya:2019uiw,Kamiya:2021hdb}, while the $p\Omega^-$ CF~\cite{Morita:2016auo} has been studied using a phenomenological Gaussian potential with a Yukawa tail fitted to lattice QCD results~\cite{HALQCD:2014okw}.

From scattering theory, the long-range Coulomb interaction modifies the asymptotic wave functions, so the standard Lippmann--Schwinger equation (LSE) is no longer applicable in its usual form. The asymptotic states are Coulomb-distorted rather than free spherical waves, expressed via regular and irregular Coulomb functions. In Ref.\,\cite{Lednicky:1981su}, R.\,Lednicky and V.\,L.\,Lyuboshits treated the Coulomb interaction exactly in the $pp$ wave function, while the strong interaction was incorporated through the asymptotic behavior of the $S$-wave at large distances. The calculation of the CF is improved by employing the effective range expansion (ERE) (see, for instance, Sec.~3.2 of Ref.~\cite{Albaladejo:2025kuv}), which effectively separates the long-range electromagnetic contributions from the short-range strong interaction. The work of Ref.~\cite{Albaladejo:2025kuv} carefully addressed the interplay between both interactions in femtoscopy, emphasizing short-distance effects that remain relevant even for extended sources and are not properly taken into account within the Lednicky-Lyuboshits (LL) approximation. The formalism includes finite-range potential (FRP) corrections, further improved by a phenomenological parameter $\beta$ fitted to the low-momentum CF.

 In Refs.~\cite{Torres-Rincon:2023qll,Encarnacion:2024jge,Xie:2026hpp}, the Coulomb interaction was included in a momentum-space chiral framework by Fourier transforming the Coulomb potential, $1/r$, after truncating it at a finite radius, $R_C \sim 60~\mathrm{fm}$. This method was applied to Goldstone boson--$D$-meson and $S=-1$ meson--baryon systems. However, the Vincent--Phatak (VP) method~\cite{Vincent:1974zz} allows a totally consistent treatment of Coulomb and strong interactions in momentum space by introducing a matching radius of the order of the (short) range of the strong potential that separates an inner region where both interactions are present from an outer Coulomb-dominated region. Matching the wave functions and their derivatives at this boundary yields exact Coulomb-modified scattering amplitudes and phase shifts. This approach has been applied to the $\Sigma^+ p$ and $\Lambda_c^+ p$ CFs~\cite{Haidenbauer:2021zvr,Haidenbauer:2020kwo} and to baryon--baryon systems with $S=-2$~\cite{Liu:2022nec}.

The present study has a twofold objective. First, we provide predictions for the $S=-4$ pseudoscalar--decuplet sector, namely the $\olsi{K}\Omega$ channel, whose correlation function can be tested in future measurements and may help constrain the repulsive mechanisms predicted by chiral dynamics. Second, we investigate the single-channel $K^-\Omega^-$ system as a clean laboratory for studying the interplay between strong and Coulomb interactions, employing the VP method to extract scattering observables within a fully consistent framework.

In addition, we propose a modification of the original VP method. In the formulation of Ref.~\cite{Vincent:1974zz}, the inner and outer regions are separated by a sharp step function. While formally well defined, this choice can lead to numerical instabilities and technical complications when solving the LSE in momentum space, due to the highly oscillatory behavior induced by the Fourier transform of the truncated Coulomb potential. Here, we replace the sharp separation by a smooth Woods--Saxon (WS) switching function, which preserves the formal structure of the method while enabling a more stable and robust numerical solution of the LSE.

This paper is organized as follows. Section~\ref{sec:II} introduces the theoretical framework for the CF and the $\olsi{K}\Omega$ strong interaction. Section~\ref{sec:III} presents the asymptotic form of the wave function and its use, together with the FRP correction, to derive an approximate expression for the CF. Section~\ref{sec:IV} describes the calculation of the exact wave function and the implementation of the modified VP treatment of the Coulomb interaction in momentum space. This framework is then used in Sec.~\ref{sec:results} to determine the scattering parameters and CFs for the $\olsi{K}\Omega$ system. Finally, Appendix~\ref{appendix} discusses the WS truncation of the long-range Coulomb interaction and demonstrates its performance not only for the $\olsi{K}\Omega$ system considered in this work, but also for the well-known proton--proton system, using the high-precision local Argonne V18 potential, which is more naturally solved in coordinate space.

\section{Correlation function and strong interaction formalism}
\label{sec:II}

The CF for a pair of hadrons is given by the 
Koonin--Pratt (KP) formula (see \textit{e.g.} \cite{Bauer:1992ffu, Lisa:2005dd,Ohnishi:2016elb,Fabbietti:2020bfg,Vidana:2023olz, Albaladejo:2024lam}),
\begin{equation}
    C(\vec k) = \int d^3 \vec r \,S(\vec r)\, |\Psi(\vec k,\vec r)|^2,
    \label{eq:KPformula}
\end{equation}
where $\vec k$ denotes the relative three-momentum of the pair. The source function $S(\vec r)$ describes the probability distribution for emitting the two particles with relative separation $\vec r$ and is commonly parametrized by a spherically symmetric Gaussian profile, $S(\vec r)=S(r)=(4\pi R_{\text{src}}^2)^{-3/2} \exp(-r^2/4R_{\text{src}}^2)$, where $R_{\text{src}}$ is the source radius. Lastly, $\Psi(\vec k,\vec r)$ is the elastic relative scattering wave function of the pair, accounting for both strong and Coulomb interactions. If the strong interaction is restricted to the $\ell=0$ partial wave, as in the present case, only the $S$-wave component of the Coulomb wave function is affected. 
The total wave function can then be decomposed as:\footnote{We follow here the conventions and normalizations of Ref.~\cite{Albaladejo:2025kuv}, where $\langle \vec p\,' | \vec p\,\rangle= (2\pi)^3\delta^3(\vec p - \vec p\,')$, similarly $\langle \vec r\,' | \vec r\,\rangle= \delta^3(\vec r - \vec r\,')$, and hence $\langle\vec r\, | \vec p\,\rangle =e^{i\vec p\cdot\vec r}$.}
\begin{subequations}\begin{align}
\Psi(\vec k,\vec r) &  = \Psi_C(\vec k,\vec r) - \psi_C^{0}(k,r) + \psi^0(k,r)\,,\\
\!\Psi_C(\vec k,\vec r) &  =  e^{-\frac{\pi\eta}{2}} \Gamma(1\!-\!i\eta)\, M(i\eta,1;-ikr\!-\!i\vec k \cdot \vec r)\,e^{i\vec k \cdot \vec r} \,,
    \label{eq:coulombanalyticwf}
\end{align}\end{subequations}
where $\Psi_C(\vec k,\vec r)$ is the full Coulomb wave function, $\psi_C^0(k,r)$ is its $\ell=0$ component, and $\psi^0(k,r)$ is the corresponding strong-interaction wave function in the presence of the Coulomb interaction. Here, $\eta = Z_1 Z_2 \alpha \mu /k$, where $\mu$ denotes the reduced mass of the hadron pair, $Z_{1,2}$ are the corresponding electric charges expressed in units of the proton charge, and $\alpha \simeq 1/137$ is the fine-structure constant. The function $M(a,b;z)={}_1F_1(a,b;z)$ represents the confluent hypergeometric function. Inserting the above decomposition into Eq.\,\eqref{eq:KPformula}, the KP formula reads~\cite{Albaladejo:2025kuv}
\begin{equation}
    C(k) = C_C(k) +\! \int \!\!d^3r\,S(r)\left( |\psi^0(k,r)|^2 \!-\! |\psi_C^0(k,r)|^2 \right),
    \label{eq:swaveKP}
\end{equation}
where $C_C(k)=\int d^3\vec r\,S(r)\,|\Psi_C(\vec k,\vec r)|^2$ is the CF generated solely by the Coulomb interaction. 

The lowest order chiral Lagrangian describing the interaction between Goldstone bosons and the baryons of the decuplet, written in terms of the decuplet spin-field tensor and the covariant derivative as defined in Refs.~\cite{Jenkins:1991es,Sarkar:2004jh}, reads
\begin{equation}
\mathcal L = -i \overline{T}{}^\mu_{abc} \gamma^\nu \mathcal D_\nu T_\mu^{abc}.
\end{equation}
As shown in Ref.~\cite{Sarkar:2004jh}, in the static limit
the $S$-wave component of the interaction kernel reduces to the WT-like contact term
\begin{equation}
    \mathcal V_{ij}(p',p;\sqrt{s}) = -\frac{C_{ij}}{4f^2}(p^0+p'^0) \, ,
    \label{eq:WTpotential1}
\end{equation}
where $p^0$ and $p'^0$ are the incoming and outgoing meson energies, respectively, the indices $i,j$ run over all possible coupled channels and $f$ denotes the pion decay constant, normalized to $f \simeq 92\,\text{MeV}$.
 In the present work, we focus on the strangeness $S=-4$ sector, for which the only accessible channel is $\olsi{K} \Omega$. Since the corresponding WT coupling is $C=-3$, the strong interaction is repulsive and prevents the formation of bound states. Together with the absence of coupled channels, this makes the system particularly well suited to isolate and test Coulomb effects.

In the non-relativistic limit, the $\olsi{K} \Omega$ interaction kernel can be expanded as
\begin{equation}
    \mathcal V(p',p) = -\frac{C}{4f^2} \left(2m_K+ \frac{p^2}{2m_K} + \frac{p'^2}{2m_K} \right) \, ,
    \label{eq:WTpotential}
\end{equation}
where $m_K$ is the kaon mass and $p$ ($p'$) the center-of-mass (CM) momentum of the incoming (outgoing) meson, respectively.

\section{Asymptotic wave-function approach and effective-range--improved LL approximation for the CF}
\label{sec:III}

The asymptotic reduced radial wave function, including both strong and Coulomb interactions, can be expressed in terms of the regular $F_0(\eta,kr)$ and irregular $G_0(\eta,kr)$ Coulomb $S$-wave functions as discussed in Refs.~\cite{Galindo:1991rxs,Albaladejo:2024lam,Albaladejo:2025kuv}:\footnote{Note that Eq.~\eqref{eq:asymptoticstrongcoulomb} follows directly from Eq.~(C.1) of Ref.~\cite{Albaladejo:2025kuv}. We note, however, that the latter contains a typographical error: the complex conjugate of the wave function $u$ was inadvertently omitted from the left-hand side.}
\begin{equation}
   [u^0_{\rm asy} (k,r)]^*\! = e^{i\sigma_0} \!\left[ \frac{F_0(\eta,\!kr)}{k} \!+\! f_{SC}(k)C_\eta^2 H^+_0(\eta,\!kr) \right].
    \label{eq:asymptoticstrongcoulomb}
\end{equation}
The radial wave function entering Eq.~\eqref{eq:swaveKP} is related to the reduced one through $\psi^0(k,r)=u^0(k,r)/r$. Here, $H_0^+(\eta,kr)=G_0(\eta,kr)+iF_0(\eta,kr)$ denotes the outgoing Coulomb wave function. Furthermore, $f_{SC}(k)$ represents the strong scattering amplitude in the presence of the Coulomb interaction, $\sigma_0=\frac{1}{2i}[\ln\Gamma(1+i\eta)-\ln\Gamma(1-i\eta)]$ is the Coulomb phase shift, and $C_\eta^2=2\pi\eta/(e^{2\pi\eta}-1)$ is the Gamow--Sommerfeld factor.

The Coulomb-modified strong scattering amplitude admits the ERE
\begin{eqnarray}
    && f_{SC}^{-1}(k) +2\eta k h^\lambda(\eta)=C_\eta^2 k (\cot\delta_{SC}-i)+2\eta k h^\lambda(\eta ) \nonumber \\
    && = -\frac{1}{a_0^{SC}}+\frac{1}{2}r_0^{SC}k^2 + \dotsc -iC_\eta^2 k\, ,
    \label{eq:fSCeffectiverangeexpansion}
\end{eqnarray}
where $\delta_{SC}$ is the strong phase shift modified by the Coulomb interaction,  and 
\begin{equation*}
    h^\lambda(\eta) = \sum_{n=1}^\infty \frac{\eta^2}{n(n^2+\eta^2)} - \ln(\lambda\eta) - \gamma_E,
\end{equation*}
with $\lambda\equiv \mathrm{sgn}[Z_1Z_2]$, and $\gamma_E\simeq0.577$ is the Euler--Mascheroni constant.

Note that Eq.\,\eqref{eq:swaveKP} requires the exact wave function $\psi^0(k,r)$ rather than its asymptotic form. When the CF $C(k)$ is evaluated by inserting the asymptotic wave function into Eq.~(\ref{eq:swaveKP}), it formally corresponds to the LL approximation, $C_{\rm LL}(k)$. Accordingly, we decompose the CF as:
\begin{subequations}
\begin{align}
    C(k) \!&=\! C_\text{LL}(k)+\delta C(k)\,,\\
   \!\!\!\!\!C_\text{LL}(k) \!&=\! C_C(k) \! + \!\!  \int \!\!\! d^3 r S(r)\big( |\psi^0_\text{asy}(k,\!r)|^2 \! - |\psi_C^0(k,\!r)|^2 \big),\\
    \delta C(k) \!&=\!  \int d^3 \! r\, S(r) \big(|\psi^0(k,r)|^2 - |\psi^0_\text{asy}(k,r)|^2 \big)\,.
\end{align}
\label{eq:CLLplusdeltaC}
\end{subequations}
The term $\delta C$ is commonly referred to as the finite-range potential (FRP) correction \cite{Lednicky:1981su,ExHIC:2017smd,Ohnishi:2016elb,Albaladejo:2024lam}, as it accounts for the effects arising from the finite spatial range of the interaction. This contribution can be evaluated by exploiting that the solution of the Schr\"odinger equation for
a local, energy-independent, and finite-range strong potential, in the presence of the Coulomb
interaction satisfies\,\cite{Preston:1993}:
\begin{align}\label{eq:fullFRP}
     & \int d^3 r \big(|\psi^0(k,r)|^2 - |\psi^0_{\rm asy}(k,r)|^2 \big) = \nonumber \\
    & -4\pi C_\eta^2 |f_{SC}(k)|^2  \frac{d}{dk^2}\Big[ \text{Re}f_{SC}^{-1}(k) + 2\eta k h^\lambda (\eta) \Big]\,. 
\end{align}
As shown in Ref.~\cite{Albaladejo:2025kuv}, Eq.~(\ref{eq:fullFRP}) leads to the following approximate expression for the FRP correction:
\begin{eqnarray}
    \delta C(k) &\simeq & -4\pi \beta \,S(0)\, C_\eta^2 |f_{SC}(k)|^2 \times \nonumber \\
    &&\frac{d}{dk^2} \left( \text{Re}f_{SC}^{-1}(k) + 2\eta k h^\lambda (\eta) \right).
    \label{eq:FRPcorrectionformula}
\end{eqnarray}
The phenomenological parameter $\beta$ is introduced in Ref.\,\cite{Albaladejo:2025kuv} to compensate for the approximation $S(r)\simeq S(0)$ over the integration region where the asymptotic wave function deviates from the exact solution. For sufficiently large sources, one expects $\beta \simeq 1$. In addition, $\beta$ may effectively account for contributions arising from the energy dependence and/or nonlocality of the short-range strong interaction that are not included in Eq.~\eqref{eq:fullFRP}. Consequently, depending on the nature of the interaction, $\beta$ may differ substantially from unity and may even become negative; see, for instance, some simple examples in Appendix B of Ref.~\cite{Albaladejo:2025kuv}. It is therefore treated as a free parameter of the model.
\section{Exact wave function}
\label{sec:IV}

The relative wave function appearing in Eq.~(\ref{eq:swaveKP}), $\psi^0(k,r)$, can be obtained from the half-off-shell scattering amplitude $T(k,q)$ (transition amplitude from an off-shell hadron-pair with momentum $q$ to the on-shell one with momentum $k$) by solving the LSE~\cite{Albaladejo:2025kuv,Vidana:2023olz}
\begin{equation}\label{eq:psi0numerical}
    [\psi^0(k,r)]^* = j_0(kr) + \frac{\mu}{\pi^2}\!\! \int_0^\infty \!\! q^2 dq  \frac{j_0(qr) T(k,q) }{k^2-q^2+i\epsilon}\,,
\end{equation}
where $j_0(kr)$ denotes the $\ell=0$ spherical Bessel function. The $S$-wave scattering amplitude entering in the above equation satisfies
\begin{equation}
    T(p',p)  =  V(p',p) + \frac{\mu}{\pi^2}\!\! \int_0^\infty \!\! q^2 dq\frac{V(q,p)T(p',q)}{k^2-q^2+i\epsilon}.
    \label{eq:numericalLS}
\end{equation}
 The potential $V(p',p)=V_S(p',p)+V_C(p',p)$ combines the strong interaction contribution together with the Coulomb interaction. As mentioned above, the long-range Coulomb interaction makes the standard momentum-space LSE formulation intractable; we therefore introduce a modified version in Subsec.\,\ref{sec:coulomb}. 

\subsection{Strong potential}

The strong component of the interaction is determined from Eq.\,\eqref{eq:WTpotential}. In order to regularize its ultraviolet behavior, we introduce the form factor
\begin{equation}
    V_S(p',p) = \frac{1}{2m_K} \mathcal V(p',p) e^{ -\frac{p^2-k^2}{\Lambda^2} } e^{ -\frac{p'^2-k^2}{\Lambda^2} }\,\,.
    \label{eq:strongpotentialoff-shell}
\end{equation}
Here, $k$ is the on-shell incoming and outgoing momenta. The potential $\mathcal V(p',p)$ is obtained from Eq.\,\eqref{eq:WTpotential} by allowing $p$ and $p'$ to be off-shell.\footnote{Chiral potentials are usually written in terms of four-momenta. If one aims to express them in terms of three-momenta, a prescription for the time components (energies) is required. Here we choose them to satisfy the dispersion relation $p^0=m_K+\vec{p}^{\,2}/2m_K$ with $\vec p$ off-shell.} The prefactor $M_\Omega/(2\mu\sqrt{s})\simeq 1/(2m_K)$ is introduced to match, in the non-relativistic limit, the normalization of the scattering amplitude obtained by solving the LSE in Eq.\,\eqref{eq:numericalLS} to that adopted in Ref.~\cite{Sarkar:2004jh}. In particular, the relation between the on-shell elastic amplitude obtained in the present work and the corresponding quantum-mechanical (QM) scattering amplitude and the phase shift $\delta(k)$ is
\begin{equation}
    T(k) = -\frac{2\pi}{\mu}f^{\rm QM}(k)
    = -\frac{2\pi}{\mu}\,\frac{e^{2i\delta(k)}-1}{2ik}\, .
\end{equation}

The Gaussian cutoff employed to regularize the off-shell part of the potential is taken as $\Lambda = 800 \pm 100$~MeV. This value reproduces the same scattering length as the sharp cutoff $\Lambda = 814$~MeV used in Ref.~\cite{Lin:2026ypf} to generate the $\Omega(2012)^-$ in the $S=-3$ sector, employing the same strong Lagrangian as in the present work. Since our study focuses on the $S=-4$ sector, we assess the sensitivity of our results to the cutoff by varying it in the range $700$--$900$~MeV. The resulting uncertainty band is included in our predictions to account for possible differences associated with the change in strangeness sector.

\subsection{Coulomb potential}
\label{sec:coulomb}

To incorporate the Coulomb interaction in momentum space, the long-range part of the potential must be regularized. To this end, we closely follow the VP approach of Ref.~\cite{Vincent:1974zz}. Although originally developed to extract strong-interaction phase shifts in the presence of Coulomb effects, the method can be straightforwardly extended to obtain the exact wave function. We define an auxiliary potential, $V_{\rm short}(r)$,  constructed from the Coulomb interaction and the original strong potential,
\begin{equation}
   V_\text{short} (r) = V_S(r) + V_C(r) g(r)\,.\label{eq:vshortvlong} 
\end{equation}
Here, $g(r)$ is a regularizing function satisfying $g(r<R_{\rm M})\simeq 1$ and $g(r\to\infty)=0$, where $R_{\rm M}$ is chosen such that the strong potential has effectively vanished for $r\geq R_{\rm M}$. In practice, $R_{\rm M}$ should be taken as small as possible while ensuring that $V_S(r)$ is negligible at larger distances. The suppression of $g(r)$ at large distances regularizes the long-range Coulomb tail, making the momentum-space LSE for $V_{\rm short}(r)$ numerically tractable. 

For $r < R_{\rm M}$, the wave function $\psi_{\rm short}$ obtained from $V_{\rm short}$ coincides, up to an overall normalization constant, with the exact solution of the combined strong-plus-Coulomb problem. This normalization is determined by matching $\psi_{\rm short}$ to the correct asymptotic strong-plus-Coulomb wave function at $r = R_{\rm M}$.

Originally, Ref.~\cite{Vincent:1974zz} proposed $g(r)=\Theta(R_M-r)$. In momentum space, however, this truncated Coulomb potential produces an oscillatory kernel which leads to numerical instabilities in the LSE. Instead, as discussed in detail in Appendix~\ref{appendix}, we adopt a WS-type regularization 
\begin{equation}
    g_{\rm WS}(r)=\frac{1+\exp(-R_{\rm WS}/b)}{1+\exp((r-R_{\rm WS})/b)}\, .
\end{equation}
The $\ell=0$ projection of the regulated Coulomb interaction is obtained from Eq.\,\eqref{eq:coulombswaveintegralgeneral}:
\begin{equation}
    \!\!\! V_C^{\ell=0}(p',p) \! = \! \frac{4\pi Z_1 Z_2 \alpha}{pp'} \!\!\! \int_0^{\!\infty} \!\! \frac{dr}{r} g_{\rm WS}(r) \sin(pr) \sin(p'r),
    \label{eq:woodssaxonprojected}
\end{equation}
which must be evaluated numerically. On the mass shell, this regulated potential exhibits a left-hand cut (LHC) below threshold for $k=i\kappa$ with $\kappa \ge 1/(2b)$, and the WS--VP method is therefore applicable only for $k^2$ above the branch point at $-1/(4b^2)$. The parameter $R_{\rm WS}$ must be chosen sufficiently large so that a matching radius $R_{\rm M}$ exists where the strong interaction has effectively vanished while $g_{\rm WS}(R_{\rm M})\simeq 1$. Conversely, $b$ must be taken sufficiently small so that the LHC remains well separated from the energy region of interest. We employ $R_{\rm WS}=5~\mathrm{fm}$ and $b=0.2~\mathrm{fm}$, for which the LHC is located at $\kappa \gtrsim 500~\mathrm{MeV}$. Additionally, we choose $R_M\sim2.5-4$ fm and check numerical stability of the procedure within that range.

By solving the LSE with the short-range potential $V_{\rm short}$, the corresponding reduced wave function of the two hadrons, $u^0_{\rm short}(k,r)$, can be obtained from Eq.~(\ref{eq:psi0numerical}). This wave function coincides, up to a normalization constant $A(k)$, with the exact strong-plus-Coulomb solution for $r<R_{\rm M}$, since the Schr\"odinger equation is the same in the inner region. At $r=R_M$, where the WS regulator has not yet deviated significantly from unity, the wave function must match the exact asymptotic form $[u^0_{\rm asy}(k,r)]^*$ given in Eq.~(\ref{eq:asymptoticstrongcoulomb}). Matching at $r=R_{\rm M}$ yields a system of two equations for the two unknowns, $A(k)$ and the exact strong amplitude $f_{SC}(k)$ in the presence of Coulomb interaction,
\begin{subequations}\label{eq:vincentPhataksystem}
\begin{align}
    A(k)\, [u^0_{\rm short}(k,r=R_{\rm M})]^* & = [u^0_{\rm asy}(k,r=R_{\rm M})]^*\,, \\
    A(k)\, \frac{d[u^0_{\rm short}(k,r)]^*}{dr}\Bigg|_{r=R_{\rm M}}  & = \frac{d[u^0_{\rm asy}(k,r)]^*}{dr}\Bigg|_{r=R_{\rm M}}\,.
\end{align}
\end{subequations}
Particular attention is devoted to the normalization constant $A(k)$, which is required to restore the correct long-distance behavior of the Coulomb interaction. This aspect has received little attention in the previous literature, and its explicit treatment in the present work constitutes a novel feature of our approach. From the system of equations above, we obtain
\begin{subequations}
\begin{align}
    A(k) &= \displaystyle\frac{ e^{i\sigma_0} }{
    \big[ H_0^+(\eta,kr), u^0_{\rm short}(k,r)^* \big]_{R_{\rm M}}}\,,
    \label{eq:coulombscreenedwavefunctionnormalization} \\
    f_{SC}^{-1}(k) &= -C_\eta^2 k\left(i +\frac{[G_0(\eta,kr),u^0_{\rm short}(k,r)^*]_{R_{\rm M}}}{[F_0(\eta,kr),u^0_{\rm short}(k,r)^*]_{R_{\rm M}}}\right)\ ,
    \label{eq:coulombscreenedwavefunctionfsc}
\end{align}
\end{subequations}
where
\begin{equation}
\big[{\cal F}(r),{\cal G}(r)\big]_{R_{\rm M}}=\left({\cal F} (r)\frac{d{\cal G}(r)}{dr}-{\cal G}(r)\frac{d{\cal F}(r)}{dr}\right)_{r=R_{\rm M}}\,\!\!\!\!\!\!,
\end{equation}
is the Wronskian of the functions ${\cal F}$ and ${\cal G}$ evaluated at $r=R_{\rm M}$, and we have used that $[G_0(\eta,kr),F_0(\eta,kr)]_{R_{\rm M}}=k$. Note that the scattering amplitude of Eq.~(\ref{eq:coulombscreenedwavefunctionfsc}) is manifestly unitary since the ratio of the second term cancels the momentum-dependent phase of $u^0_{\rm short}$, leading to ${\rm Im} [f_{SC}^{-1}]=-C_\eta^2 k$. The Coulomb-modified strong phase shift, $\delta_{SC}$, can be extracted from the scattering amplitude $f_{SC}(k)$ [see Eq.\,\eqref{eq:fSCeffectiverangeexpansion}].

The exact reduced radial wave function, which is the solution of the strong plus Coulomb interaction $(V_S + Z_1Z_2\alpha/r)$, can then be written as:\footnote{For relative distances larger than $R_{\rm M}$, the exact wave function $[u^0(k,r)]^*$ starts to deviate significantly from $A(k)\,[u^0_{\rm short}(k,r)]^*$ once these distances are similar to $R_{\rm WS}$. This is not an issue, since the short-range potential $V_S$ has already vanished and the wave function has entered the asymptotic regime, fully determined by the Coulomb-modified strong scattering amplitude $f_{SC}(k)$ obtained above. Choosing $R_{\rm M}$ as small as possible improves numerical stability. In practice, the method is robust and any residual dependence on the auxiliary parameters $R_{\rm M}$, $R_{\rm WS}$ and $b$ can be reliably controlled.}
\begin{equation}\label{eq:exact}
\!\!\![u^0(k,r)]^* \!=\! \left\{
\begin{array}{@{}ll@{}}
A(k)\,[u^0_{\rm short}(k,r)]^*\,, & r \! \le \! R_{\rm M},\\[1.5ex]
\begin{aligned}[b]
e^{i\sigma_0}\Big[ & F_0(\eta,kr)/k \\
& + f_{SC}(k)\,C_\eta^2\,H_0^+(\eta,kr)\Big]\,,
\end{aligned} & r \! > \! R_{\rm M}\,.
\end{array} \right.
\end{equation}
As a final remark, when the Coulomb potential is truncated by a step function, $g(r)=\Theta(R_M-r)$, Eq.~\eqref{eq:coulombscreenedwavefunctionfsc} readily reproduces Eq.~(8) of Ref.~\cite{Vincent:1974zz}. In this case, $u^0_{\rm short}(k,r=R_M)$ reduces to a linear combination of $\sin(kR_M)$ and $\cos(kR_M)$ (see Eq.~(7) of Ref.~\cite{Vincent:1974zz}). More generally, however, Eq.~\eqref{eq:exact} provides the exact wave function for any regulator $g(r)$ of the long-range tail of the Coulomb potential.

\section{Results}\label{sec:results}
With the modified VP formalism introduced above, we show in Appendix~\ref{appendix} that, for an analytically solvable potential, the WS--VP method reproduces results for the strong-plus-Coulomb scattering amplitude $f_{SC}$, both below and above threshold. 

\begin{figure*}[t]
    \centering
    \includegraphics[width=0.99\linewidth]{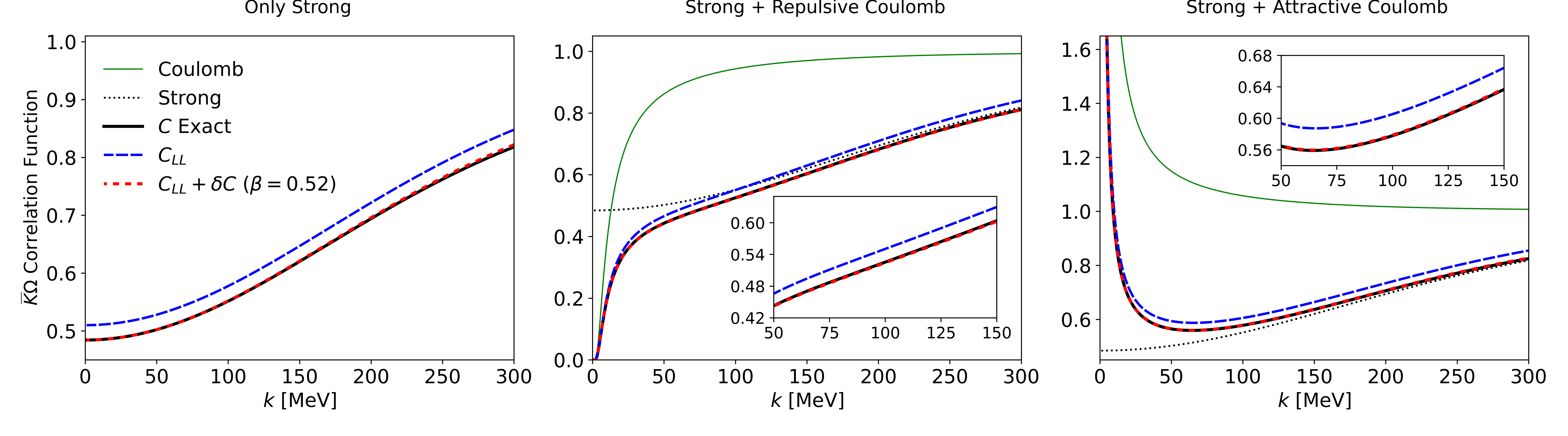}
    \caption{$\olsi{K}\Omega$ CF for $\Lambda=800$ MeV and $R_{\text{src}}=0.5$ fm in three different scenarios: strong interaction only (left panel), strong interaction in the presence of a repulsive Coulomb interaction (middle panel), and strong interaction in the presence of an attractive Coulomb interaction (right panel). In each case, we compare the CF obtained from the exact wave function, $C(k)$, with that calculated from the asymptotic wave function using the extracted scattering amplitude $f_{SC}$, denoted by $C_{\rm LL}(k)$. The approximate FRP correction of Eq.~\eqref{eq:FRPcorrectionformula}, with $\beta=0.52$, is added to $C_{\rm LL}$, resulting in the dashed lines. The insets illustrate the impact of this correction in the intermediate-$k$ region. Finally, the green curves in the middle and right panels correspond to $C_C(k)$, the CF generated by the Coulomb interaction alone.\label{fig:CF_academic}}
\end{figure*}

\begin{table}[t]
    \centering
    \begin{tabular}{ccc}  \hline 
        \tstrut & $a^{SC}_0$ [fm] & $r^{SC}_0$ [fm] \\ \hline  
       \tstrut Only Strong & $0.293$ & $0.509$ \\
        Strong + Repulsive Coulomb & $0.306$ & $0.538$ \\
        Strong + Attractive Coulomb & $0.282$ & $0.493$ \\ \hline
    \end{tabular}
    \caption{Effective range expansion parameters (Eq.~\eqref{eq:fSCeffectiverangeexpansion}) of the $\olsi{K}\Omega$ system for $\Lambda=800$~MeV in three scenarios: the strong interaction only, the strong interaction supplemented by a repulsive Coulomb potential, and the strong interaction supplemented by an attractive Coulomb potential.}
    \label{tab:scatteringparametersLambda800}
\end{table}

The same VP method can be applied to the off-shell potential of Eq.~(\ref{eq:strongpotentialoff-shell}) to extract $f_{SC}$ and obtain the asymptotic wave function of Eq.~(\ref{eq:asymptoticstrongcoulomb}), as well as the exact wave-function at short distances $A(k)[u^0_{\rm short}(k,r)]^*$ of the first line of Eq.~\eqref{eq:exact}. We compare: 
\begin{enumerate}[topsep=1pt,itemsep=-1ex,partopsep=1pt,parsep=5pt,labelindent=0pt,itemindent=1em,leftmargin=15pt]
    \item The $\olsi{K} \Omega$ CF computed from the exact wave function obtained by solving the Lippmann--Schwinger equation in momentum space, with the Coulomb interaction incorporated through the WS--VP scheme.
    \item The $\olsi{K} \Omega$ CF computed with the asymptotic wave function, using $f_{SC}$ extracted from the exact solution. This CF corresponds to $C_{\rm LL}$ in Eq.~(\ref{eq:CLLplusdeltaC}), which can be corrected by $\delta C$ to better reproduce the exact CF for an appropriate choice of the $\beta$ parameter.
\end{enumerate}

We study the $\olsi{K}\Omega$ system by first considering only the off-shell strong interaction of Eq.~\eqref{eq:strongpotentialoff-shell}. We then include both repulsive and attractive Coulomb contributions on top of the strong potential. The attractive Coulomb case, unphysical for $\olsi{K}\Omega$, is presented here for illustrative, if academic, purposes. Finally, we present theoretical predictions, including uncertainty bands, for the physically relevant $\olsi{K}\Omega$ channels: $\olsi{K}{}^0\Omega^-$, which is free from Coulomb effects, and $K^-\Omega^-$, where the Coulomb interaction is repulsive. In practice, the $\olsi{K}^0\Omega^-$ channel is not easily accessible experimentally, since reconstructing the $K^0_L$ component is challenging. In addition to $C_{K^-\Omega^-}$, the most likely observable to be measured is therefore $C_{K^0_S\Omega^-}=(C_{K^0\Omega^-}+C_{\olsi{K}^0\Omega^-})/2$. A similar treatment is done in the work of Ref.\,\cite{Albaladejo:2025lhn} for the $\pi^{\pm} K^0_S$ CF. Our calculation provides a prediction for only one of the two components entering the $K^0_S\Omega^-$ CF. A meaningful comparison with experimental data would thus require a prediction for the other component, which belongs to the $S=-2$ sector and lies beyond the scope of the present work. The main purpose of this work is instead to demonstrate the performance of the VP method for treating Coulomb interactions in momentum space.

Figure~\ref{fig:CF_academic} shows the $\olsi{K}\Omega$ CF for the three scenarios considered: strong interaction only (left panel), strong interaction supplemented by a repulsive Coulomb potential (middle panel), and strong interaction supplemented by an attractive Coulomb potential (right panel). In all cases, the calculations are performed using the off-shell strong potential of Eq.~(\ref{eq:strongpotentialoff-shell}) with $\Lambda=800$~MeV. The source radius is chosen to be $R_{\rm src}=0.5$ fm so that the effect of the FRP correction is more visible. The corresponding scattering parameters for the three cases are compiled in Table~\ref{tab:scatteringparametersLambda800}. Several features can be appreciated in the figure. First, the implementation of the Coulomb interaction, and in particular the normalization factor $A(k)$, reproduces the vanishing (diverging) behavior at low momenta for the repulsive (attractive) case. This behavior arises because, at low momenta, $C(k)\propto C_\eta^2$ as discussed in Ref.~\cite{Albaladejo:2025kuv}; without the proper normalization of the wave functions, this limit would not be recovered. In fact, if one were to naively solve the regularized Coulomb potential without properly normalizing the wave functions, the resulting CF would neither vanish nor diverge in the correct limits. Secondly, at low momenta the strong-plus-Coulomb correlation function is largely dominated by the Coulomb interaction and closely follows the pure Coulomb result. As the momentum increases, Coulomb effects progressively diminish and the CF approaches the purely strong-interaction limit.                                

The CF extracted from the asymptotic wave function, $C_{\rm LL}$, differs from the actual CF due to the short-distance behavior of the wave function, which is governed by the renormalized half-off-shell scattering amplitude. The latter may contain ambiguities (see, e.g., Appendix A.6 of Ref.~\cite{Nieves:1999bx}) that are inherited by theoretical calculations of the CF and, within the KP formalism, must be compensated by an appropriate choice of the source describing the emission of the interacting hadron pair~\cite{Epelbaum:2025aan}.

Although the $C_{\rm LL}$ approximation works very well in the particular case of the repulsive strong potential given in Eq.\,\eqref{eq:strongpotentialoff-shell}, as illustrated in Fig.\,\ref{fig:CF_academic}, the corrections can become significantly larger for other off-shell extrapolations, interaction models, or less extended spatial sources. Such differences within the KP framework should be accounted for through the FRP correction, $\delta C$, introduced in Eqs.\,\eqref{eq:CLLplusdeltaC}. The approximate expression for the FRP correction in Eq.\,\eqref{eq:FRPcorrectionformula} is proportional to $|f(k)|^2 r_0$ at low momenta and includes a parameter $\beta$, which effectively absorbs effects associated with the source finite size, as well as the energy dependence and non-localities of the interaction potential, which are closely tied to its off-shell behavior. In general, the smallness of the FRP correction does not necessarily imply a small effective range $r^{SC}_0$ [see Table~\ref{tab:scatteringparametersLambda800} for the strong interaction of Eq.~\eqref{eq:strongpotentialoff-shell}], but rather may reflect a partial cancellation among the different contributions entering the difference between $C$ and $C_{\rm LL}$. This cancellation is effectively encoded in the value $\beta = 0.52$ for the case shown in Fig.\,\ref{fig:CF_academic} and we see that it provides an excellent reproduction of the exact $C(k)$ CF (the black solid and red-dashed curves are almost indistinguishable in the plots). In the present study, $\beta$ is fine-tuned such that the FRP correction reproduces, at threshold in the Coulomb-less case, the difference between $C_{\rm LL}$ and the exact CF. The same value is then used for all three cases, where it also provides an excellent description of the exact CFs. We stress that this value of $\beta$ is neither universal nor predictive, but rather a phenomenological parameter encoding effects associated with the particular off-shell extension adopted here. Different off-shell extensions yielding the same on-shell amplitude may therefore lead to different values of $\beta$. 

In practice, in the analysis of measured CFs, a simultaneous fit of the phenomenological $\beta$ parameter together with the on-shell amplitude may provide an effective way to account for off-shell ambiguities in femtoscopic CFs.\footnote{Reference~\cite{Molina:2025lzw} proposes a simultaneous fit of the on-shell amplitude and the source radius.} In this regard, the results shown in Figs.~2--4 of Ref.~\cite{Albaladejo:2025kuv} support this strategy.

\begin{figure}[t]
    \centering
    \includegraphics[width=0.95\linewidth]{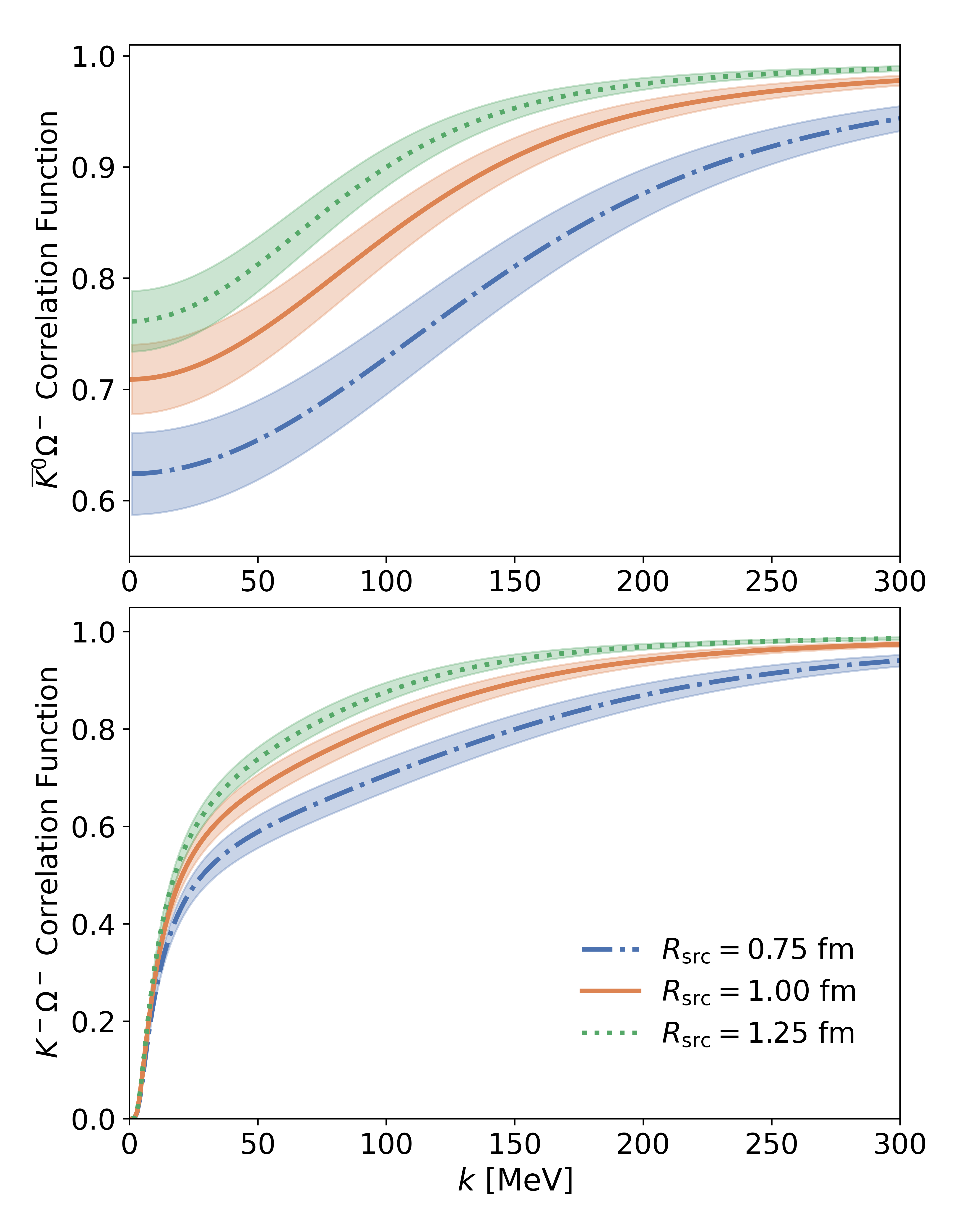}
    \caption{$\olsi{K}{}^0 \Omega^-$ and $K^-\Omega^-$ CFs for three Gaussian source sizes. The results are obtained using the strong interaction of Eq.~\eqref{eq:strongpotentialoff-shell}, and including Coulomb repulsion for the $K^-\Omega^-$ system via the WS-extended VP method. The shaded bands represent the uncertainties obtained by propagating those of the cutoff $\Lambda$ and the source size $R_{\text{src}}$. See the main text for further details.}
    \label{fig:CF_physical}
\end{figure}

We now present in Fig.~\ref{fig:CF_physical} the theoretical predictions for $K^-\Omega^-$, which is directly accessible experimentally, together with those for the $\olsi{K}^0\Omega^-$ system. The latter is not directly accessible, but constitutes one of the two components entering the measurable $K^0_S\Omega^-$ CF, as discussed above. For clarity, only the CFs calculated with the exact wave functions are shown. The shaded bands represent the 68\% confidence intervals obtained by Monte Carlo propagating the uncertainties associated with both the cutoff parameter $\Lambda$ and the source size $R_{\text{src}}$. Specifically, we adopt $\Lambda = 800 \pm 100$~MeV and consider source radii $R_{\text{src}} = 0.75$, $1.0$, and $1.25$~fm, each with a relative uncertainty of 10\%. In each system, the CFs exhibit similar shapes and differ primarily in their overall strength, which is largely determined by the source size. Note that the CF is below unity, as expected for a repulsive interaction. At low momenta, the width of the band is dominated by the uncertainty in $\Lambda$, while at higher momenta the source size becomes more relevant. Experimental data could, in principle, help discriminate the source size and constrain the cutoff in different momentum regions for the $\olsi{K}{}^0\Omega^-$ system. For the $K^-\Omega^-$ system the Coulomb interaction masks the low momenta region and one has to rely on higher momenta. 

To conclude this section of results, we also propagate the uncertainty associated with $\Lambda$ to the scattering parameters of the $\olsi{K}{}^0\Omega^-$ and $K^-\Omega^-$ systems, and find
\begin{subequations}\begin{align}
\!a_0^{\olsi{K}{}^0 \Omega^-}         & \!\!= 0.293\pm0.031\,\text{fm}\,,%
&\!\!\! r_0^{\olsi{K}{}^0 \Omega^-} & \!\!= 0.509^{+0.019}_{-0.001}\,\text{fm}\,,\\
\!a_0^{K^- \Omega^-}                  & \!\!= 0.306\pm0.034\,\text{fm}\,,%
&\!\!\! r_0^{K^- \Omega^-}          & \!\!= 0.538^{+0.019}_{-0.001}\,\text{fm}\,.
\end{align}\end{subequations}
The qualitative difference between the neutral and charged channels is preserved, demonstrating that the influence of Coulomb interactions is largely insensitive to variations in the cutoff $\Lambda$.

\section{Conclusions}\label{sec:conclusions}

We have developed a momentum-space framework for the simultaneous treatment of strong and Coulomb interactions in femtoscopy. 

We have shown that the VP method, extended by means of a WS regulator for the long-range tail of the Coulomb potential, achieves improved numerical stability while providing a robust and reliable implementation of Coulomb effects in the LSE at the short distances relevant for femtoscopic observables. Furthermore, it enables the extraction of strong-interaction phase shifts in the presence of Coulomb forces.

We have investigated the role of the asymptotic wave function in the calculation of CFs and quantified its deviations from the exact result through the FRP correction. We have found that these deviations can be effectively encoded in a phenomenological parameter, $\beta$, which may absorb effects associated with both the finite size of the source and the off-shell structure of the interaction, the latter being one of the main obstacles to extracting detailed information on hadron--hadron interactions from femtoscopic measurements in a model-independent way.

The formalism has been applied to the $\olsi{K}\Omega$ system, which provides a particularly clean setting to study the interplay between strong and Coulomb interactions. As an additional validation, we have also considered the well-studied $pp$ system using the high-precision Argonne V18 nucleon--nucleon potential. The corresponding results are presented in Appendix~\ref{appendix}.

Finally, we have presented theoretical predictions for the physically relevant $\olsi{K}{}^0\Omega^-$ and $K^-\Omega^-$ systems, including uncertainty estimates associated with the ultraviolet cutoff and the source size. Future experimental measurements of these CFs could provide valuable constraints on the interactions between antikaons and the lowest-lying decuplet baryons.

Moreover, the framework developed in this work is sufficiently general to enable straightforward extensions to more complex systems, including coupled-channel dynamics, which require the calculation of wave functions both above threshold for open channels and below threshold for closed channels.

\begin{acknowledgments}
This work is part of the Grants {\small PID2023-147458NB-C21} and {\small CEX2023-001292-S} funded by {\small MICIU/AEI/10.13039/501100011033} and by ERDF/EU, as well as of the Grant {\small CIPROM/2023/59} funded by Generalitat Valenciana {\small 10.13039/501100003359}. %
M.\,A.\,acknwoledges the \guillemotleft{}Ramón y Cajal\guillemotright{} program Grant {\small RYC2022-038524-I} funded by {\small MICIU/AEI/10.13039/501100011033} and by ESF+, and the \guillemotleft{}Atracción de Talento\guillemotright{} program Grant {\small PIE 20245AT019} funded by {\small CSIC 10.13039/501100003339}. %
M.\,A. and A.\,F. warmly thank the support from ACVJLI. I.V. acknowledges in particular D. Unkel for the many discussions held during the course of this work.
\end{acknowledgments}

\appendix

\section{Regularized Coulomb interaction}\label{appendix}

In this Appendix, we discuss the choice of the regularizing function $g(r)$ entering the auxiliary short-range interaction defined in Eq.~(\ref{eq:vshortvlong}). It is required to satisfy two basic conditions: the Coulomb interaction should remain unaffected at short distances, $g(r<R_{\rm M})\simeq 1$, while being smoothly suppressed at large distances, such that $g(r\to\infty)=0$. In momentum space, the $\ell=0$ projection of the regulated Coulomb interaction, $Z_1Z_2 \alpha g(r)/r$, reads
\begin{align}
   \!\!\!V_C^{\ell=0}(p',p) & = \frac{1}{2} \! \int_{-1}^{+1} \!\!\!\!\!\!\! d \cos\Theta_{pp'} \!\! \int \!\! d^3\!r \frac{Z_1Z_2 \alpha}{r} g(r) e^{i\vec{r}\cdot(\vec{p}-\vec{p
    }^{\,\prime})} \nonumber \\
    & =  \frac{4\pi Z_1\! Z_2 \alpha}{pp'}\!\! \int_0^\infty \! \!\frac{dr}{r}g(r)\sin(pr)\sin(p'\!r) .
    \label{eq:coulombswaveintegralgeneral}
\end{align}
For positive momentum magnitudes $p'$ and $p$, convergence of Eq.~(\ref{eq:coulombswaveintegralgeneral}) requires that $g(r)$ decreases faster than $1/r$ at large distances. Originally, Ref.~\cite{Vincent:1974zz} uses $g(r)=\Theta(R-r)$, for which the integral in Eq.~(\ref{eq:coulombswaveintegralgeneral}) can be evaluated analytically, yielding 
\begin{eqnarray}
    V_C^{\ell=0}(p',p) = \frac{2\pi Z_1 Z_2 \alpha}{pp'} \bigg[ && \!\!\!\!\!\! {\rm Ci}[(p-p')R] - {\rm Ci}[(p+p')R] \nonumber \\
    &&+ \log\frac{(p+p')R}{(p-p')R}  \bigg]\,, \label{eq:Theta}
\end{eqnarray}
where ${\rm Ci}[x]=-\int_x^\infty \frac{\cos t}{t}\,dt=\gamma_E + \ln x + \sum_{n=1}^\infty \frac{(-1)^n x^{2n}}{2n(2n)!}$ is the cosine integral. This potential can be analytically continued when either $p$, $p'$, or both are purely imaginary,\footnote{Indeed, Eq.~(\ref{eq:coulombswaveintegralgeneral}) can also be evaluated directly; thanks to the step function, the integral converges for any complex values of $p$ and $p'$.} as required for solving the LSE below threshold, since no LHC discontinuity appears. However, for positive momenta it becomes highly oscillatory, leading to numerical instabilities in the solution of the LSE. This behavior worsens notably as the coordinate-space cutoff $R$ increases. A simple alternative is a Yukawa-screened Coulomb interaction, $g(r)=\exp(-m_\gamma r)$, where the screening mass $m_\gamma$ is chosen sufficiently small so that the Coulomb interaction remains essentially unchanged at distances relevant to the strong interaction. In this case, Eq.~(\ref{eq:coulombswaveintegralgeneral}) can also be solved analytically, yielding
\begin{eqnarray}
    V_C^{\ell=0}(p',p) = \frac{Z_1Z_2 \pi \alpha}{pp'}\log\frac{(p+p')^2+m_\gamma^2}{(p-p')^2 + m_\gamma^2}\, .
    \label{eq:yukawaprojected}
\end{eqnarray}
The momentum-space dependencies are smooth and, more importantly, lack any oscillatory behavior, which facilitates the solution of the LSE above threshold. 

The situation is more restrictive below threshold (negative energies). In this regime, the on-shell momentum becomes imaginary, and convergence of Eq.~(\ref{eq:coulombswaveintegralgeneral}) requires that $g(r)\exp[{\rm Im}(p+p')r]$ vanishes faster than $1/r$ as $r \to \infty$. In particular, when computing $V_C^{\ell=0}(i\kappa,i\kappa')$ with the Yukawa regulator, the integral in Eq.~(\ref{eq:coulombswaveintegralgeneral}) diverges when $\kappa+\kappa' \ge m_\gamma$. On the diagonal $\kappa=\kappa'$, this condition reduces to $\kappa=m_\gamma/2$, corresponding to the well-known Yukawa LHC branch point, which also constrains the analytic continuation of Eq.~(\ref{eq:yukawaprojected}) to imaginary momenta. Taking into account that $m_\gamma$ must be chosen sufficiently small so that the Yukawa exponential remains close to unity over the range of the strong interaction, the applicability of this screened Coulomb interaction below threshold is therefore very limited.

Ensuring convergence generally requires either that $g(r)$ decreases faster than exponentially (super-exponential decay) or that it has compact support, that is, $g(r)=0$ for $r>R$. Although super-exponential regulators such as $g(r)=\exp(-a r^n)$ with $n>1$, or $g(r)=1-{\rm erf}(ar)$, satisfy the required convergence properties, they typically induce an exponential growth of $V_C^{\ell=0}(p',p)$ for imaginary momenta, leading to significant numerical instabilities. Regulators with compact support avoid this issue, but tend to generate oscillatory structures in momentum space, similar to those produced by the sharp step-function choice $g(r)=\Theta(R-r)$.

To balance convergence, numerical stability, and smoothness, we employ a WS-type regularization of the Heaviside step function,
\begin{equation}
    g(r)=\frac{1+\exp(-R_{\rm WS}/b)}{1+\exp((r-R_{\rm WS})/b)}
\end{equation}
Unlike the previous cases, the momentum-space Coulomb interaction is obtained by numerically evaluating the integral in Eq.~(\ref{eq:coulombswaveintegralgeneral}). Although the WS regulator exhibits an asymptotic exponential decay, $g(r)\sim\exp(-r/b)$, and therefore does not guarantee convergence for arbitrary complex momenta, it provides a practical compromise between physical fidelity and numerical stability. In particular, the parameter $R_{\rm WS}$ determines the distance up to which the Coulomb interaction remains essentially unmodified, while $b$ controls the smoothness of the transition to the asymptotic region. Moreover, the regulator-induced LHC is located at $\kappa=1/(2b)$ and can therefore be shifted arbitrarily far from the region of interest by choosing $b$ sufficiently small. In the limit $b\to 0$, the regulator approaches $g(r)=\Theta(R_{\rm WS}-r)$; consequently, $b$ controls both the location of the LHC and, together with $R_{\rm WS}$, the degree of oscillatory behavior of the resulting momentum-space potential.

\begin{figure}
    \centering
    \includegraphics[width=0.8\linewidth]{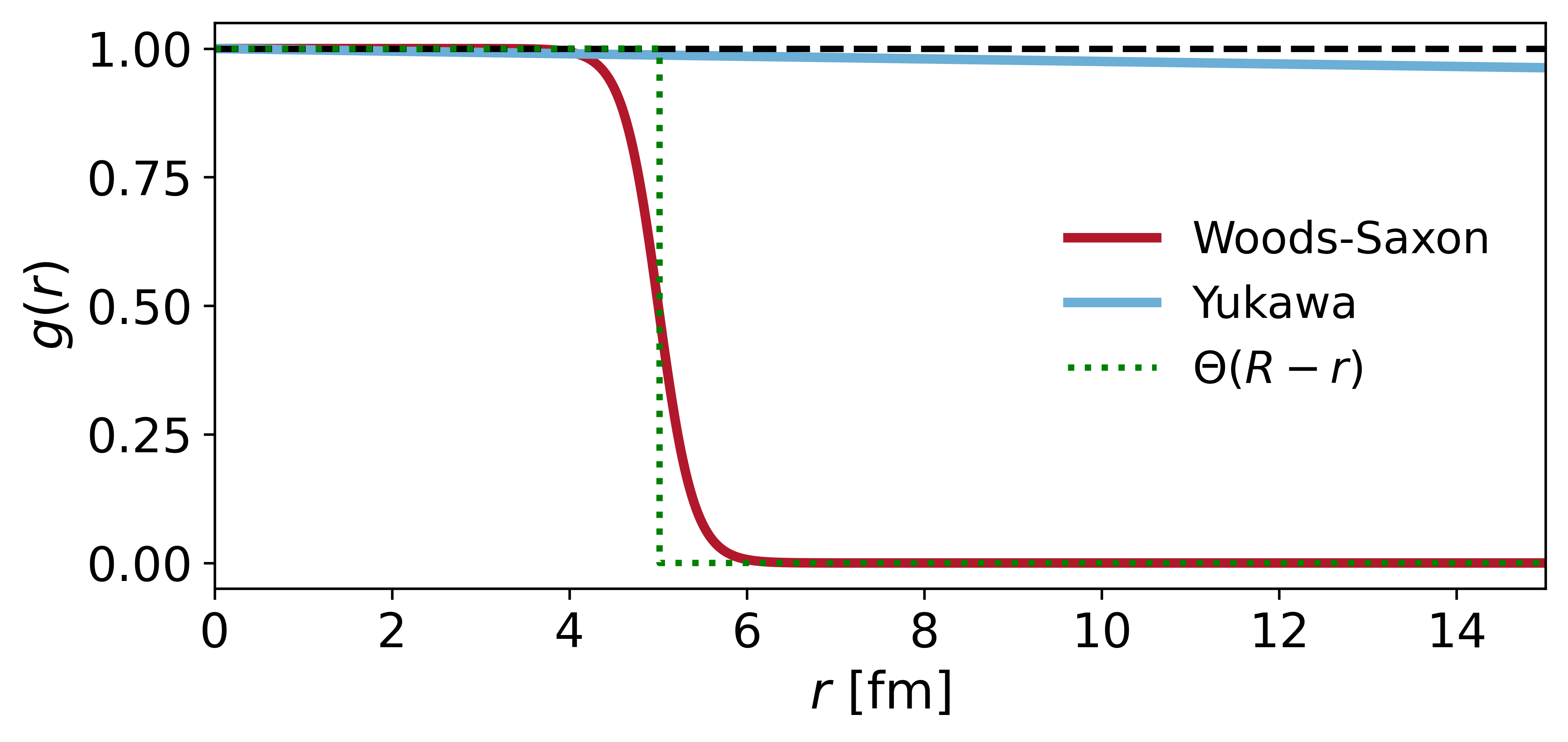}
    \caption{Yukawa (with $m_\gamma = 0.5\,\mathrm{MeV}$), WS (with $R_{\rm WS} = 5\,\mathrm{fm}$ and $b = 0.2\,\mathrm{fm}$), and sharp step-function (with $R = 5\,\mathrm{fm}$) form factors, $g(r)$, shown in blue, red, and green, respectively.}
    \label{fig:YWS}
\end{figure}
In the following, we examine the two Coulomb-screening schemes introduced above: the Yukawa and WS regulators. In the comparisons below, we employ $m_\gamma=0.5$~MeV for the Yukawa potential, and $R_{\rm WS}=5$~fm and $b=0.2$~fm for the WS regulator. These parameter choices ensure deviations of no more than $1\%$ from the pure Coulomb interaction at $r\sim 4$~fm, as shown in Fig.~\ref{fig:YWS}. Beyond this distance, however, the WS regulator rapidly approaches zero, whereas the Yukawa regulator remains much closer to unity, deviating by less than $5\%$ even at $r=20$~fm. The matching radius $R_M$ is chosen in the range $2.5-4$ fm for stability.

We employ the on-shell-factorized $\olsi{K} \Omega$ potential constructed from Eq.~(\ref{eq:strongpotentialoff-shell}):
\begin{equation}
  V^{\rm onF}_S(p,p')=\frac{1}{2m_K}\mathcal V(k)
  e^{-\frac{p^2-k^2}{\Lambda^2}}
  e^{-\frac{p'^2-k^2}{\Lambda^2}}, \label{eq:pot-onshell}
\end{equation}
for $\Lambda=800$~MeV. For this interaction, the LSE admits the following exact solution~\cite{Albaladejo:2025kuv}
\begin{equation}
   f_{SC}(k)= -\frac{\mu}{2\pi}\frac{T_{SC}(k)}{C_\eta^2e^{2i\sigma_0}}
    =-\frac{\mu}{2\pi}\frac{1}
   {[V^{\rm onF}_S(k)]^{-1}-J_0(k)}\,,
    \label{eq:definitionscoulombfSC}
\end{equation}
where the Coulomb-dressed loop function is
\begin{eqnarray}
    J_0(k) &=&2\mu \int \frac{d^3 q}{(2\pi)^3}
    \frac{2\pi\eta_q}{e^{2\pi\eta_q}-1}
    \frac{[f_\Lambda(q,k)]^2}
    {k^2- q^2+i\epsilon }
    \nonumber\\
    &+&\Theta(-\lambda)\,
    \frac{\alpha\mu^2}{\pi}
    \Big[\psi(i\eta)+\psi(-i\eta)+2\gamma_E  \Big]\,, \label{eq:j0loop-dressed}
\end{eqnarray}
with $\psi(z)$ the digamma function and $\eta_q=Z_1Z_2\mu\alpha/|\vec q|$. The form factor is inherited from the potential,
\begin{equation}
    f_\Lambda(q,k)=\exp\!\left[-\frac{q^2-k^2}{\Lambda^2}\right].
\end{equation}
We note that, in the attractive case, the contribution proportional to
$(\psi(i\eta)+\psi(-i\eta))$, arising from the Coulomb bound states in the second term of the Coulomb-dressed loop function, diverges in the $k\to 0$ limit. This divergence exactly cancels the infrared divergence generated by the integration of the first contribution to Eq.~\eqref{eq:j0loop-dressed}. This exactly solvable model provides a convenient benchmark for assessing the accuracy of the extended VP method with both the Yukawa and WS regulators, as the resulting amplitudes can be directly compared with the exact result, Eq.~\eqref{eq:definitionscoulombfSC}.

\begin{figure}[t]\centering
\includegraphics[width=\linewidth]{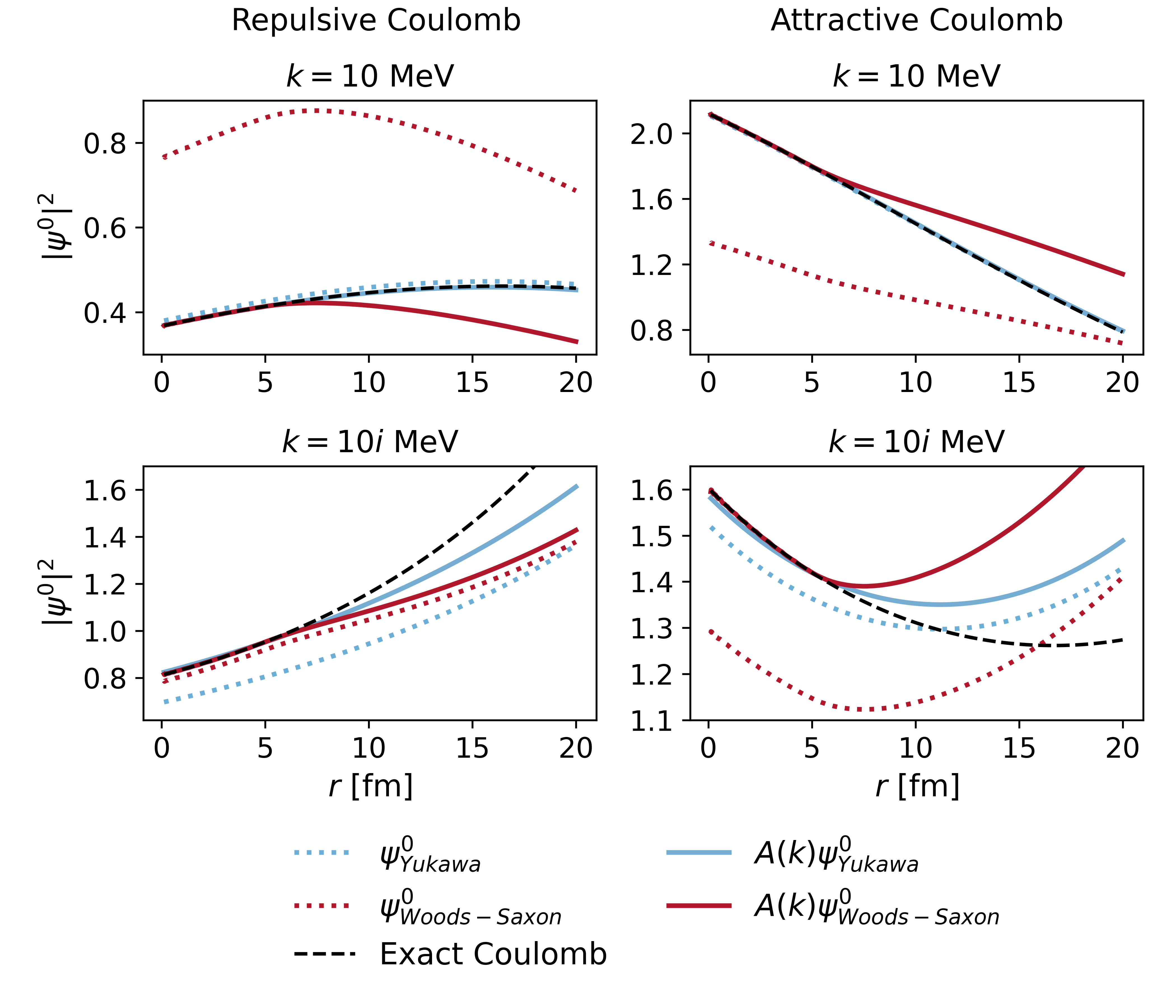}
\caption{Comparison, as a function of $r$, between the modulus squared of the analytic $\ell=0$ Coulomb wave function (black dashed lines), $|e^{i\sigma_0}F_0(\eta,kr)/kr|^2$,
and the corresponding wave-function moduli squared obtained from Coulomb potentials with Yukawa (blue) and WS (red) regularizations. The top panels correspond to a real momentum of $10\,\mathrm{MeV}$ (above threshold), while the bottom panels correspond to an imaginary momentum of $10i\,\mathrm{MeV}$ (below threshold). The left and right panels show the repulsive and attractive Coulomb cases, respectively. Solid (dashed) lines correspond to results obtained with (without) the VP normalization constant $A(k)$, which, in the absence of the strong interaction, is given by Eq.~\eqref{eq:AsinCb}.\label{fig:CoulombWF}}
\end{figure}

First, we set $V^{\rm onF}_S(p,p')=0$. Figure~\ref{fig:CoulombWF} compares the exact Coulomb wave function, $|\psi_C^0(k,r)|^2$, with those obtained using the Yukawa- and WS-regulated potentials. In the absence of the strong interaction, Eqs.\,\eqref{eq:vincentPhataksystem}, matching at $R_M=0$ for convenience, reduce to:
\begin{subequations}\begin{align}
A(k) & = \frac{e^{i\sigma_0}C_\eta}{[\psi_{\rm short}^0(k,0)]^*}\,.\label{eq:AsinCb}\\
f_{SC}(k) & = 0\,.
\end{align}\end{subequations}
Results are presented both above and below threshold for attractive and repulsive Coulomb interactions. Above threshold, both regulators reproduce the exact solution with similar accuracy within their respective domains of applicability (see Fig.~\ref{fig:YWS}), especially in the physically relevant region $r \leq 4$~fm for the VP method including the strong potential. We emphasize that the normalization constant $A(k)$ is essential for reproducing the correct wave function.

\begin{figure}[t]\centering
\includegraphics[width=\linewidth]{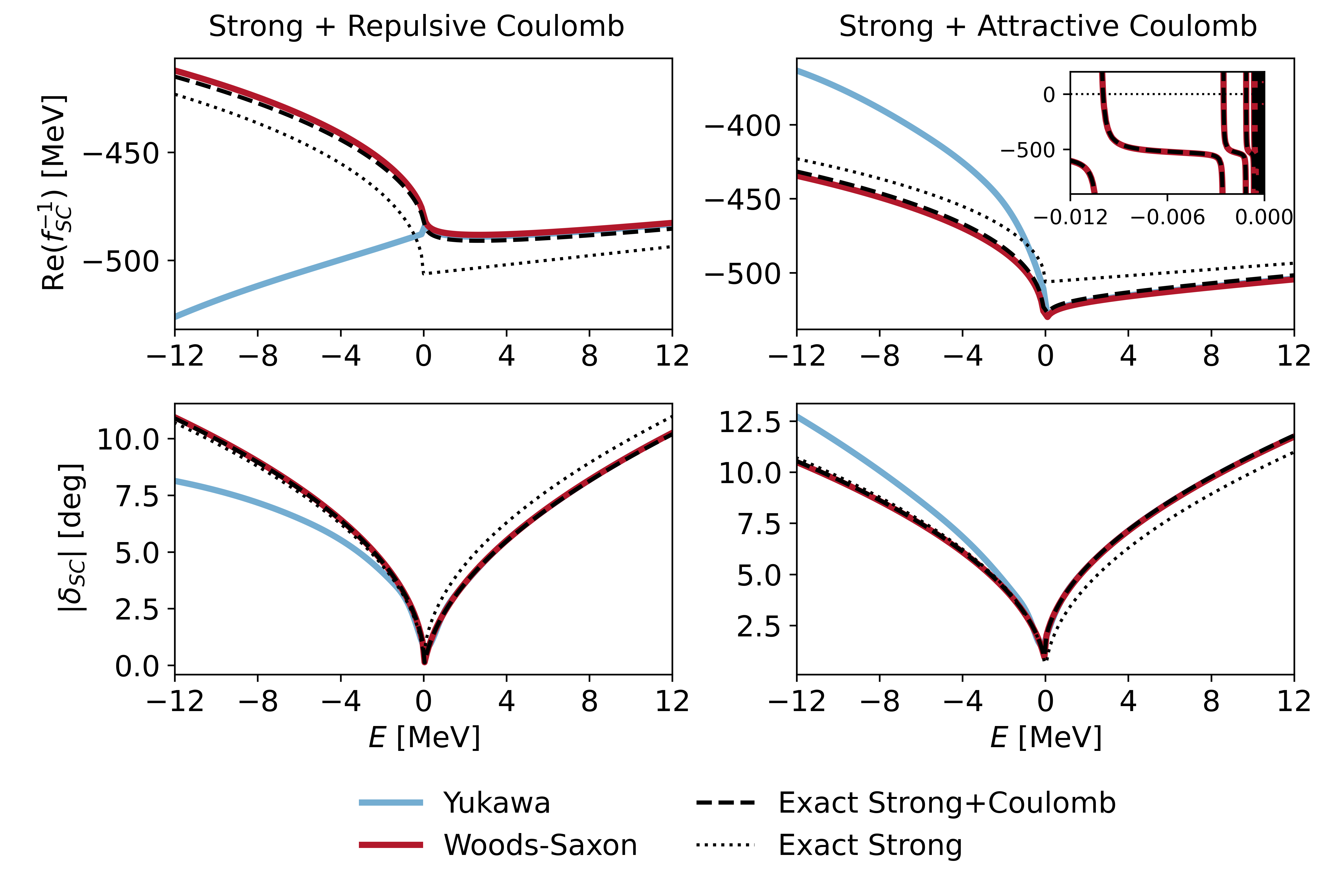}
\caption{Real part of $f_{SC}^{-1}(k)$ (top panels) and modulus of the strong phase shift in the presence of the Coulomb interaction, $|\delta_{SC}|$ (bottom panels), as functions of the energy below and above threshold, obtained using the on-shell-factorized strong potential of Eq.~\eqref{eq:pot-onshell}. The modified VP method is implemented with the Yukawa (blue curves) and WS (red curves) momentum-space regulators. Results for repulsive and attractive Coulomb interactions are shown in the left and right panels, respectively. The exact solution obtained from Eq.~\eqref{eq:definitionscoulombfSC} is shown by the dashed black curve, while the dotted black curve represents the corresponding solution in the absence of the Coulomb interaction. The structure of the results at subthreshold energies around $-\mu (Z_1 Z_2 \alpha)^2/2 \simeq -0.01~\mathrm{MeV}$ is not clearly visible in the figure; however, our calculation captures the intricate features of the Coulomb dynamics in this region, including the emergence of Coulomb bound states shifted by the strong interaction in the attractive case, as can be inferred from the inset.
\label{fig:CoulombTSCComparison}}
\end{figure}

In the subthreshold region, the WS wave functions perform well for $r \leq 4\,\text{fm}$, whereas the Yukawa parametrizations fail to reproduce the correct $r$-dependence, despite the normalization at the origin being enforced. This discrepancy becomes evident in the $10$--$20\,\text{fm}$ region, even though at $r = 20\,\text{fm}$ the Yukawa-screened potential differs from the pure Coulomb interaction by only about $5\%$. This behavior is a consequence of the Yukawa LHC, which does not affect the above-threshold wave functions shown in the top panels, where the Yukawa potential reproduces the exact Coulomb wave functions accurately in the whole range displayed in the figure.

In Fig.~\ref{fig:CoulombTSCComparison} we show ${\rm Re}[f_{SC}^{-1}](k)$ (top panels) and the modulus of the strong phase shift in the presence of the Coulomb interaction $|\delta_{SC}|$ (bottom panels). These quantities are extracted using the VP method and compared with the exact solution obtained from the on-shell-factorized potential $V^{\rm onF}(p,p')$. Note that ${\rm Re}[f_{SC}^{-1}]$ fully determines the amplitude $f_{SC}$, since ${\rm Im}[f_{SC}^{-1}]$ is zero for negative energies and $-C_\eta^2 k$ for positive energies. For energies below threshold\footnote{For negative energies, the wave function is non-normalizable except at the discrete energies corresponding to bound states, which are determined by the condition $\cot\delta_{SC} = i$. Note that below threshold, $\delta_{SC}$ is generally complex because of the Coulomb interaction. However, the inverse amplitude, $
f^{-1}_{SC} = C_\eta^2\, k (\cot\delta_{SC} - i)$, is real, as it must be.} the phase shift is determined through the analytic continuation of $\cot\delta_{SC}= i + (C_\eta^2 k f_{SC})^{-1}$. Above threshold, the modified VP method accurately reproduces the exact result for both momentum-space regulators. Below threshold, however, the Yukawa regulator breaks down due to the presence of the Yukawa-LHC starting at $E^{\rm Yukawa}_{\rm LHC}=-m^2_\gamma/(8\mu)\simeq-80$ eV, whereas the WS regulator remains in excellent agreement with the exact solution throughout the energy range displayed in the figure, where Coulomb effects are still appreciable. In particular, the inset highlights the low-energy region where the Coulomb bound states, modified by the strong interaction, are located (i.e., the zeros of $f_{SC}^{-1}$). Note that the WS-induced LHC branch point lies far below the energy region of interest, at $E_{\rm LHC}^{\rm WS}\simeq -(8\mu b^2)^{-1}\approx -300~\mathrm{MeV}$.

\begin{figure}[t]\centering
\includegraphics[width=\linewidth]{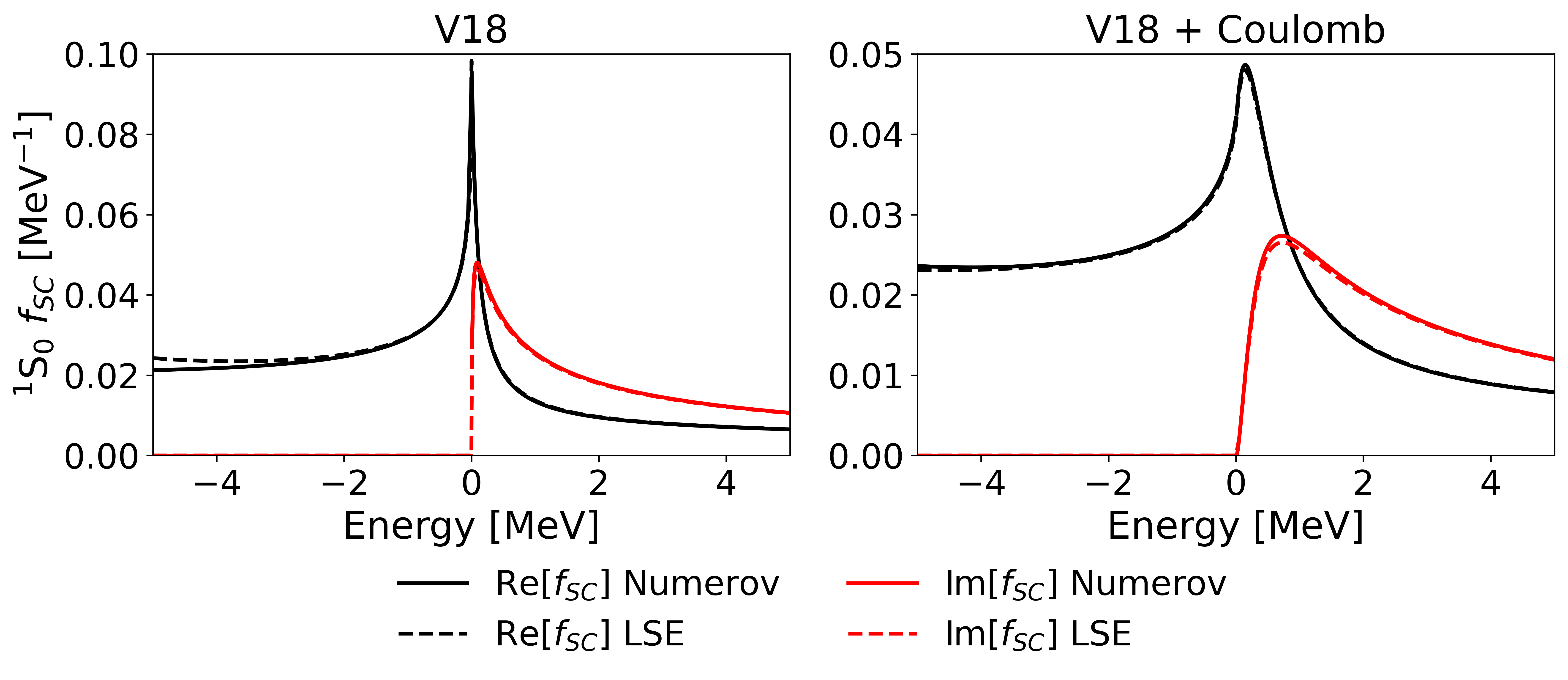}
\caption{The $^{1}S_{0}$ proton--proton scattering amplitude as a function of the CM energy. The left and right panels show the results obtained with the Argonne V18 potential without and with the Coulomb interaction, respectively. Above threshold, the solid lines represent coordinate-space solutions of the Schr\"odinger equation obtained with the Numerov method, where the Coulomb interaction is treated exactly. Below threshold, the solid lines show the analytic continuation of the ERE, Eq.~\eqref{eq:fSCeffectiverangeexpansion}. The dashed lines represent solutions of the LSE over the entire energy range, with the Coulomb interaction treated using the VP method and a WS regulator with $R_{\rm WS}=14$ fm and $b=0.2$ fm, together with a matching radius $R_M=10$ fm.
} 
\label{fig:pp_scattmat}
\end{figure}

In addition, we explore the robustness of the method with respect to the choices of $R_M$, $R_{\rm WS}$, and $b$. The VP--WS method renders the extracted amplitude $f_{SC}$ essentially independent of these auxiliary parameters, provided that $R_M$ is larger than the range of the strong interaction, and that the WS regulator remains sufficiently close to unity at $R_M$. In practice, $R_M$ should be kept as small as possible while satisfying the former condition, to facilitate the numerical calculation. Likewise, $R_{\rm WS}$ should be minimized while ensuring that the regulator remains sufficiently close to unity at $R_M$, whereas $b$ should be maximized while keeping the regulator-induced LHC at $\kappa=1/(2b)$ sufficiently far from the region of interest. To quantify the residual dependence on these parameters, Table~\ref{tab:VPWS_stability} shows the relative deviations of ${\rm Re}(f_{SC}^{-1})$ at $E=\pm1$~MeV, for a repulsive Coulomb interaction, for different choices of the WS and matching parameters. The reference values are those obtained with $R_M=3$~fm, $R_{\rm WS}=5$~fm, and $b=0.2$~fm, corresponding to the red curve in Fig.~\ref{fig:CoulombTSCComparison}. We find that the relative deviation between the different choices of parameters is negligible and thus the extracted $f_{SC}$ can be considered independent of this choice. Furthermore, the relative deviation of all these results from the analytical result also shown in Fig.~\ref{fig:CoulombTSCComparison} is approximately $0.5\%$, which can reasonably be attributed to the numerical accuracy of the calculation, e.g., in solving the LSE or performing the Fourier transform of the WS-truncated Coulomb potential, rather than to the choice of auxiliary parameters.
\begin{table}[t]
\centering
\begin{tabular}{cccccc}
\hline\hline
$R_M$ [fm] & $R_{\rm WS}$ [fm] & $b$ [fm]
& $1-g_{\rm WS}(R_M)$
& $\Delta_{+}$ 
& $\Delta_{-}$
\\
\hline
$2$ & $3$ & $0.1$ & $0.005\%$   & $0.00056\%$ & $0.00033\%$ \\
$2$ & $5$ & $0.2$ & $0.00003\%$ & $0.00037\%$ & $0.00109\%$ \\
$4$ & $6$ & $0.2$ & $0.005\%$   & $0.00012\%$ & $0.00204\%$ \\
$4$ & $7$ & $0.3$ & $0.005\%$   & $0.00028\%$ & $0.00118\%$ \\
$6$ & $7$ & $0.2$ & $0.7\%$     & $0.04022\%$ & $0.07895\%$ \\
$6$ & $8$ & $0.3$ & $0.13\%$    & $0.01279\%$ & $0.04119\%$ \\
\hline\hline
\end{tabular}
\caption{Stability of the VP--WS method above and below threshold with respect to the auxiliary parameters $R_M$, $R_{\rm WS}$, and $b$. The relative deviations of ${\rm Re}(f_{SC}^{-1})$ at $E=\pm1$~MeV, for a repulsive Coulomb interaction, are shown for different choices of these parameters. The reference values, obtained with $R_M=3$~fm, $R_{\rm WS}=5$~fm, and $b=0.2$~fm, are those shown by the red curve in Fig.~\ref{fig:CoulombTSCComparison}, and are denoted by
$y_\pm^{\rm ref}={\rm Re}(f_{SC}^{-1})|_{E=\pm1\,{\rm MeV}}$.
The corresponding relative deviations are defined as
$\Delta_\pm=\left|[y_\pm(R_M,R_{\rm WS},b)-y_\pm^{\rm ref}]/y_\pm^{\rm ref}\right|$.}
\label{tab:VPWS_stability}
\end{table}

 Finally, we further validate the formalism by applying it to the well-established proton--proton system in the $^1S_0$ partial wave. As a benchmark for our treatment of the Coulomb interaction, we first solve the Schr\"odinger equation in coordinate space using the Numerov method \cite{Numerov1924} with the Argonne V18 nucleon--nucleon potential \cite{Wiringa:1994wb}, treating the Coulomb interaction, $V_C(r)=\alpha/r$, exactly. We note that only the strong-interaction component of the Argonne V18 potential is retained in the calculation, while none of the additional electromagnetic terms included in the full V18 interaction (see Ref.~[44]) is included. The electromagnetic interaction is instead incorporated separately through the Coulomb potential. This choice exploits the locality of the Argonne V18 potential, which allows the Schr\"odinger equation to be solved accurately and efficiently in coordinate space, providing a reliable reference against which the momentum-space implementation can be assessed. We should note, however, that for negative energies, the extraction of the proton--proton scattering amplitude becomes technically challenging with the Numerov method, since the asymptotic oscillatory Coulomb solutions are replaced by a linear combination of exponentially growing and decaying functions, making the matching to the scattering boundary conditions ill-defined and numerically unstable. Figure~\ref{fig:pp_scattmat} shows, as solid lines, the real and imaginary parts of the $^1S_0$ proton--proton scattering amplitude, calculated with and without the inclusion of the Coulomb interaction using the Numerov method. Below threshold, the solid lines are obtained from the analytic continuation of the ERE, Eq.~\eqref{eq:fSCeffectiverangeexpansion}, including terms up to $\mathcal{O}(k^4)$, since the Numerov method cannot be directly applied at negative energies, as discussed above.

\begin{figure}[t]\centering
\includegraphics[width=\linewidth]{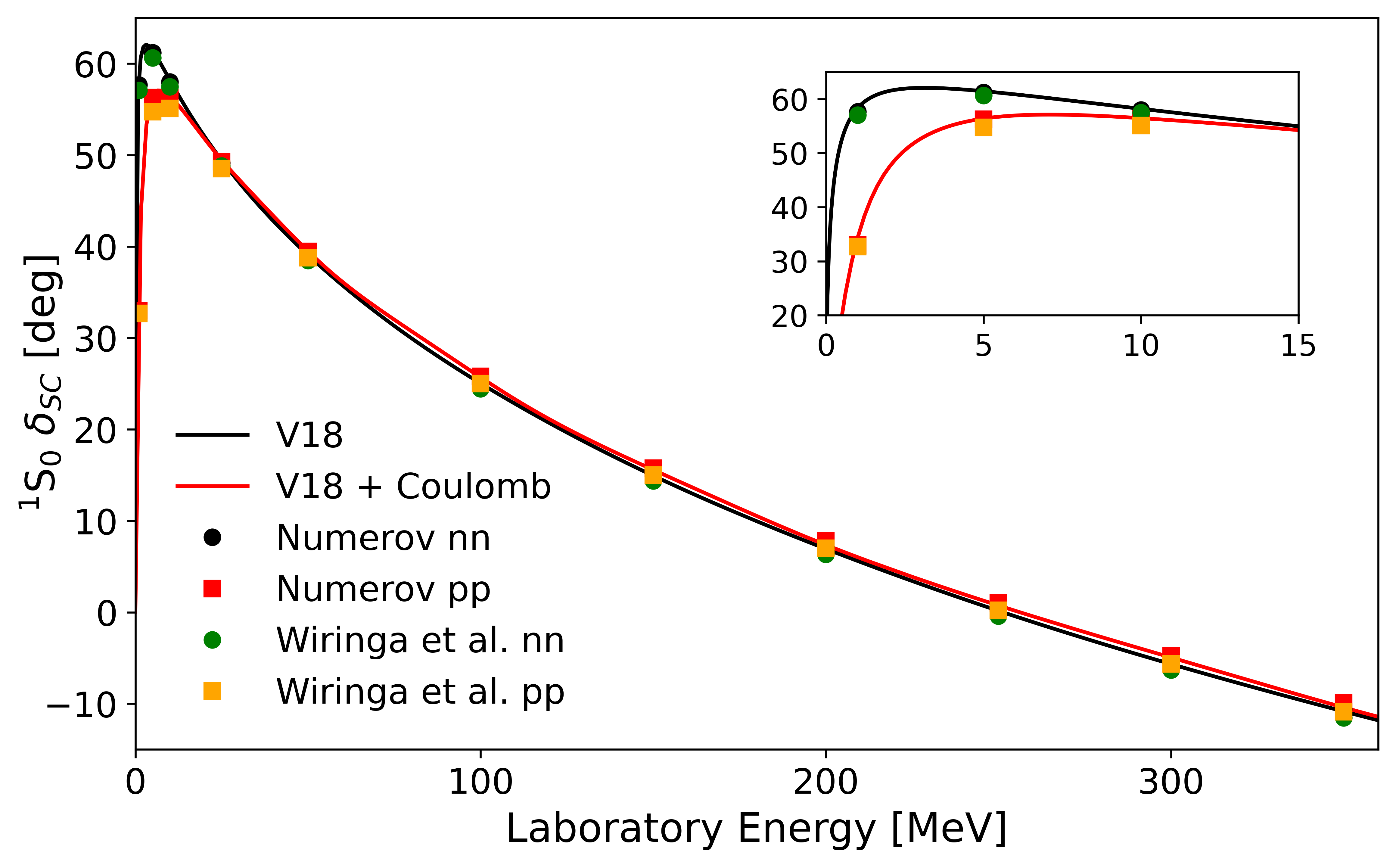}
\caption{
$^1S_0$ nucleon--nucleon phase shifts as functions of the laboratory energy ($E_{\rm lab}=2E_{\rm CM}$), obtained from the LSE using the strong terms of the Argonne V18 potential. The proton--proton results include the Coulomb interaction, treated with the modified WS--VP method. The calculated phase shifts are compared with those obtained by solving the Schr\"odinger equation in coordinate space using the Numerov method, with the Coulomb interaction treated exactly in the $pp$ channel. For completeness, we also show the phase shifts reported by Wiringa et al. in Tables~IV (neutron--neutron) and V (proton--proton) of Ref.~\cite{Wiringa:1994wb}, which include the additional electromagnetic corrections considered in the full Argonne interaction.
\label{fig:pp_ps}}
\end{figure}
We then perform the Fourier transform of the Argonne V18 potential and of the WS-truncated Coulomb interaction used in the VP method, in order to obtain their momentum-space representations required to solve the LSE. As already discussed, the VP--LSE framework can be straightforwardly extended to complex momenta (negative energies) once the interaction kernel has been specified. This allows for a controlled extension of the $f_{SC}$ scattering amplitude away from the real axis. In practice, the continuation can be performed using the same numerical solution evaluated at complex values of $k$, while consistently enforcing the correct analytic structure of the Green’s function, including the appropriate Riemann sheet.
Results for the real and imaginary parts of the $^1S_0$ proton--proton scattering amplitude $f_{SC}$, obtained by solving the LSE with the V18 and Coulomb interactions included, using the WS--VP method with $R_{\rm WS}=14$ fm and $b=0.2$ fm, are shown by the dashed lines in the right panel of Fig.~\ref{fig:pp_scattmat}. The matching radius, where the V18 nuclear potential is assumed to be negligible, is taken $R_M=10$ fm.
The excellent agreement with the results obtained from the solution of the Schr\"odinger equation above threshold provides a validation of the VP--LSE formalism introduced in this work. The dashed lines in the left panel display the Argonne V18 LSE scattering amplitude in the absence of the Coulomb interaction.

Finally, in Fig.~\ref{fig:pp_ps}, we present the $^1S_0$ proton--proton and neutron--neutron phase shifts as functions of the laboratory energy, obtained using the WS--VP method. 
The phase shifts obtained by solving the Schr\"{o}dinger equation in coordinate-space with the Numerov method treating exactly the Coulomb interaction are shown
for comparison. The good agreement with the results obtained using the WS--VP method provides further validation of  momentum-space treatment of the Coulomb interaction adopted in this work.
A comparison of the phase shifts reported in the original paper of 
Ref.\ \cite{Wiringa:1994wb}, which include the electromagnetic components of the 
 Argonne V18 interaction, are also shown for completeness.


\bibliographystyle{apsrev4-1_MOD}
\bibliography{RefsKbarOmegaCF}
\end{document}